\documentclass[twocolumn,prb,groupedaddress,showpacs,amsfonts,amssymb,amsmath,letterpaper]{revtex4-1}
\usepackage{amsmath}
\usepackage{graphicx}
\usepackage{amssymb}
\usepackage{epstopdf}
\usepackage{adjustbox}

\makeatletter

\usepackage{color}
\begin{document}

\title{
Demonstrating non-Abelian statistics of Majorana fermions using twist defects}

\author{Huaixiu Zheng}
\author{Arpit Dua}
\author{Liang Jiang}
\affiliation{\textit{Departments of Applied Physics and Physics, Yale University, New Haven, Connecticut, USA}}

\date{\today}

\begin{abstract}
We study the twist defects in the toric code model introduced by Bombin [Phys. Rev. Lett. 105, 030403 (2010)].
Using a generalized 2D Jordan-Wigner transformation, we show explicitly the twist defects carry unpaired Majorana zero modes.
We also draw the same conclusion using two alternative approaches of a perturbation theory and a projective construction.
We propose a quantum non-demolition measurement scheme of the parity of Majorana modes.
Such a scheme provides an alternative avenue to demonstrate the non-Abelian statistics of Majorana fermions with measurement-based braidings.
\end{abstract}


\maketitle

\section{Introduction}

Quantum systems exhibiting topological order have stimulated a lot of excitement in the past few decades
because a new paradigm beyond Landau's symmetry breaking theory is needed to describe those states \cite{NayakRMP08,HasanRMP10,QiRMP11}.
In particular, topology rather than symmetry should be used to characterize topological states.
The existence of topological degeneracy of ground states makes topological systems
a promising candidate for topological quantum computation (TQC) \cite{NayakRMP08,SternSci05}
as such degeneracy depends only on the topology and hence is robust against local perturbations.

One important example of topological order is the toric code model \cite{KitaevAP03,WenPRL03} which is exactly solvable.
With local interactions between spins, the system has topologically degenerate ground states with long-range entanglement.
Local perturbations only introduce exponentially small splittings of the ground state degeneracy, and the ground state subspace is thus protected from local probes and possesses a long-range topological order.
Anyons emerge as excitations in the toric code model. 
Because fusion of two such anyons has a definite outcome, these are abelian anyons and have no computational power if we want to encode quantum information in the fusion channels.

An interesting variation of the toric code model that supports non-Abelian anyons was proposed by Bombin \cite{BombinPRL10} via introducing twist defects into the model.
By verifying the fusion and braiding rules, the twists were shown to mimic the behavior of more exotic non-Abelian Ising anyons which can be used to realize the quantum gates of the Clifford group for TQC \cite{NayakRMP08}.
An important distinction was made by You and Wen \cite{YouPRB12} that those twists
are not intrinsic non-Abelian anyons themselves since they are not excitations of the Hamiltonian.
Instead, twists are extrinsic defects with projective non-Abelian statistics \cite{YouPRB12,BarkeshliPRB13}.
With a mean-field treatment, twists are shown to carry unpaired Majorana fermions \cite{YouPRB12}.
More recently, the topological entanglement entropy was calculated to show that twist defects have the same quantum dimension and fusion rules as Ising anyons \cite{BrownPRL13}.
The twists have also been proposed as qubits for quantum computation in a surface code implementation to reduce the space-time cost \cite{HastingsArXiv14}.

In this work, we \textit{explicitly} show the emergence of unpaired Majorana zero modes associated with the twist defects using three different approaches: 1) a generalized 2D Jordan-Wigner transformation \cite{Jordan_Wigner28,FradkinPRL89}; 2) a projective construction from a Majorana surface code model; 3) a perturbation theory from the Kitaev honeycomb lattice model \cite{KitaevAP06,PetrovaPRB13,PetrovaPRB14}.
In addition, we show that the parity operator of two unpaired Majorana modes is indeed the logical operator defined in the spin language, which can be used to manipulate the degenerate states of the corresponding twists. 
To demonstrate the non-Abelian statistics of twists through braiding, one needs to move the twists by changing the Hamiltonian \cite{BombinPRL10}.
Here, instead we want to realize the braiding of Majorana fermions without actually moving them.
We investigate the measurement-only topological quantum computation scheme using forced measurements \cite{BondersonPRL08,BondersonPRB13} and find that the measurement-based braiding can be simulated efficiently using a single cycle of topological charge measurements followed by a single logical qubit Pauli operation. This way, we avoid the uncertainty associate with forced measurement whose number of measurements is probabilistically determined.
Even through signatures of Majorana fermions have been identified in several spin-orbit coupled materials placed close to superconductors \cite{MourikSci12,DasNatPhys12,PergeSci14}, demonstrating the non-Abelian statistics of Majorana fermions in the condensed matter setting is still a very challenging task as it requires a combination of experimental capabilities including braiding, fusing and measuring two Majorana zero modes.
We instead show that it is possible to verify the non-Abelian statistics of Majorana fermions using the twist defects in the surface code model \cite{BravyiArXiv98,FreedmanFCM01,DennisJMP02,FowlerPRA12} on a 2D planar lattice. 
We believe all the required operations are within experimental reach in the surface code setup given recent significant experimental progress in superconducting circuits \cite{BarendsNat14,KellyNat15,CorcolesNatComm15}.
Therefore, our scheme provides an alternative approach to demonstrate the non-Abelian statistics of Majorana fermions. If supplemented with a single-qubit $\pi/8$ gate through the magic state distillation \cite{FowlerPRA12,SternSci05}, the measurement-based braiding of Majorana fermions can be used for universal quantum computation in the surface code setting \cite{HastingsArXiv14}.

Here is the outline of the paper. In Section\,\ref{MZM}, we first introduce the toric code model with twist defects and show explicitly there are unpaired Majorana fermions associated with the twists.
In Section\,\ref{Non-Abelian}, we show how to demonstrate the non-Abelian statistics of Majorana fermions using the twist defects.
In particular, we give a detailed account of the measurement-based braiding of Majorana fermions and show that it can be done in a single cycle of parity measurements.
Finally, we conclude in Section\,\ref{Conclusion}.

\section{Unpaired Majorana Zero Modes from Twist Defects}
\label{MZM}
Because of their exotic non-Abelian statistics, 
Majorana fermions \cite{Majorana37} have generated tremendous interest in the condensed matter community and a huge amount of effort has been devoted to the search of Majorana fermions in the past few years \cite{Alicea12,Leijnse12,Beenakker13,Stanescu13,ElliottRMP15,DasSarmaArXiv15}. 
So far, majority of the activity has focused on topological superconductors supporting localized zero-energy modes \cite{MourikSci12,DasNatPhys12,PergeSci14}.
However, it has been shown by \citet{PetrovaPRB13,PetrovaPRB14} that introducing topological defects in the Abelian phase of the Kitaev honeycomb model \cite{KitaevAP06} provides an alternative way to generate unpaired Majorana zero modes.
Previously, twist defects in the toric code model are shown to exhibit Ising anyon like behavior through either fusion and braiding rules \cite{YouPRB12,BrownPRL13,BarkeshliPRB13} or a mean-field treatment \cite{BombinPRL10}.
Here, it is our aim to demonstrate explicitly the emergence of unpaired Majorana zero modes associated with the twist defects as has been done in the Kitaev honeycomb model \cite{PetrovaPRB13,PetrovaPRB14}.

\begin{figure}[tb]
\centering
\includegraphics[width=0.48\textwidth]{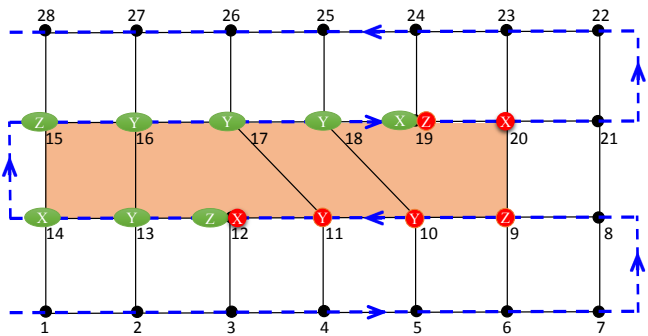}
\caption{(color online) Planar surface code lattice with an open boundary condition. The spins live on the corners of the square lattice. The dashed blue line shows the ordering of the spin along the Jordan-Wigner path. The green ovals show one possible representation of the parity operator of the two twist defects. After multipling the plaquette operators in the shaded region, the parity operator becomes the one showing in red circles.}
\label{fig:Fig1-JWT}
\end{figure}

The toric code model was initially introduced by Kitaev \cite{KitaevAP03} and later reformulated by Wen \cite{WenPRL03}.
To be self-contained, here we describe the fundamental aspects of the toric code model.
The model is defined on a 2D lattice with spin-$1/2$ particles seating on each site, see Fig.\,\ref{fig:Fig1-JWT}.
We start with the reformulated Hamiltonian with twist defects introduced by Bombin \cite{BombinPRL10}:
\begin{equation}
 H_{TCM}=-\sum_k A_k,
\end{equation}
where the plaquette operator $A_k$ is a four-body interaction term between spins living on the corners of the plaquette $k$ and is given by
\begin{equation}
\label{eq:AK1}
A_k = \raisebox{-5.3mm}{ \includegraphics[scale=0.3]{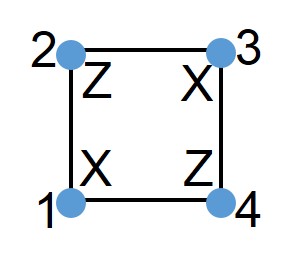} } = X_1Z_2X_3Z_4.
\end{equation}
Here, $X$'s and $Z$'s are the spin-$1/2$ Pauli operators.
For the plaquette containing a twist defect, the plaquette operator has to be modified to
\begin{equation}
\label{eq:AK2}
 A^{\prime}_k= \raisebox{-5.5mm}{ \includegraphics[scale=0.3]{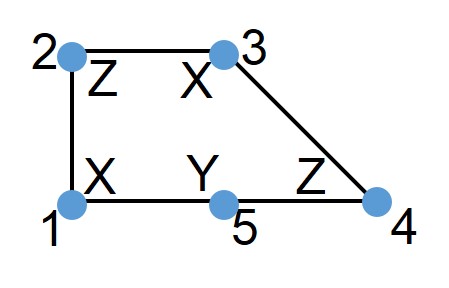} } = X_1Z_2X_3Z_4Y_5,
 \end{equation}
where $Y=iXZ$.

It is easy to check that all the plaquette operators commute with each other: each $A_k$ (or $A^{\prime}_k$) shares \textit{two} spins with its neighboring plaquettes and anti-commutation relations between different Pauli operators guarantee $A_k$ (or $A^{\prime}_k$) commutes with them; for all the other plaquettes, $A_k$ (or $A^{\prime}_k$) simply shares \textit{zero} spin with them. 
This nice property makes the toric code model exactly solvable.
Since $A^2_k=(A^{\prime}_k)^2=1$, each plaquette operator has two eigenvalues $\pm 1$.
Therefore, the ground state energy corresponds to simultaneously minimizing the contribution from each plaquette.
This leads to the condition that each $A_k$ (or $A^{\prime}_k$) takes the eigenvalue $+1$ and the ground state is a common eigenstate of all the plaquette operators: $A_k|GS\rangle=|GS\rangle$ for all $k$.
When the lattice is placed on a torus, the ground state has a degeneracy of $4\times2^{N_{\text{site}}-N_{\text{plaq}}}$ for an even$\times$even lattice, and $2\times2^{N_{\text{site}}-N_{\text{plaq}}}$ for an even$\times$odd or odd$\times$odd lattice \cite{YouPRB12}.
The factor of $4$ or $2$ is the number of distinct set of excitations allowed by the underlying lattice topology.
$N_{\text{site}}$ and $N_{\text{plaq}}$ denote the number of sites and plaquettes, respectively.
In contrast, with an open boundary condition, the ground states support gapless edge modes due to Majorana fermions \cite{WenPRL03}. 

Excitations can be generated by applying spin operators, say $Z$ on site $1$ in Eq.\,(\ref{eq:AK1}).
It flips the states of $A_k$ together with $A_k$'s diagonal neighbor that shares site $1$ with $A_k$.
Hence, excitations are localized on the plaquettes and they are always produced in pairs with a minimal energy gap of $2$.
To label the excitations, one can color the plaquettes into dark and light groups \cite{BombinPRL10}.
This labeling can be done consistently on the lattice for the regions free of twist defects.
Then a electric charge $e$ (magnetic charge $m$) is attached to excitations living in a dark (light) plaquette.
Both $e$ and $m$ can be moved by applying spin operators to flip the plaquette states, for instance, we can move the charge on $A_k$ in Eq.\,(\ref{eq:AK1}) by applying $Z$ on site 3.
Connecting those spin flipping operators gives a string operator having its ends on two plaquettes where the excitations are generated.
Once we keep moving one of the charge until the two ends of the string operator meet on the same plaquette again, the two charges will then annihilate and the system is back to the ground state.
This tells us that fusing two $e$ or $m$ charges gives vacuum $\bf{1}$.
The composition of one $e$ charge and one $m$ charge however gives a fermion $\epsilon=e\times m$.
The fermionic nature of $\epsilon$ can be verified by exchanging a composite pair of $e\times m$ with another pair of $e\times m$, which gives rise to a $\pi$-phase shift in the wave-function.
$\bf{1}$, $e$, $m$ and $\epsilon$ together forms the quantum double model of $\bf{Z}_2$ \cite{Pachos12}.
They are abelian anyons because fusing any two of them always has a definite outcome.
For an even$\times$even lattice placed on the torus, these four types of charges can be distinguished and the $4$-fold degeneracy of the ground state due to the lattice topology.
While for an even$\times$odd or odd$\times$odd lattice, $e$ becomes $m$ if moved across the boundary and hence they are no longer distinguishable any more, so are $\bf{1}$ and $\epsilon$, giving the $2$-fold topological degeneracy of the ground state \cite{YouPRB12}.

Introducing twist defects provides an alternative approach to change the ground state degeneracy.
$n$ twist defects will reduce the number of plaquette $N_{\text{plaq}}$ by $n/2$ and hence increase the degeneracy by $2^{n/2}$.
Each twist has a quantum dimension of $\sqrt{2}$ which coincides with that of Majorana fermions.
This connection to Majorana fermions has generated quite some interest \cite{BombinPRL10,YouPRB12,BarkeshliPRB13,BrownPRL13,HastingsArXiv14} mainly because of the generation of non-Abelian anyons from an Abelian phase.
Of course, caution has to be applied because twist defects are not intrinsic anyons because they are not part of the excitation spectrum of the Hamiltonian.
Instead, they are extrinsic anyons with projective non-Abelian statistics \cite{YouPRB12,BarkeshliPRB13}.
But still, twist defects have been shown to have some computational power \cite{BombinPRL10} and can be used to encode qubits for quantum computation \cite{HastingsArXiv14}.
Here, we show explicitly that twists are indeed associated with unpaired Majorana zero modes using three different techniques.

In this paper, we assume a 2D planar lattice with an open boundary condition.
Since we are only concerned with the topological degeneracy introduced by the twist defects (instead of the degeneracy associated with lattice topology), the boundary condition becomes irrelevant.
More importantly, a 2D planar lattice is more realistic from the experimental point of view. 
Hereafter, we will refer to the 2D planar model as the (planar) surface code model if we are considering a Hamiltonian or surface code if we are using a stabilizer formulation without Hamiltonian. 

\subsection{Jordan-Wigner transformation of the surface code model}
To define a Jordan-Wigner transformation on a 2D lattice, we order the spins in a quasi-1D manner as shown in Fig.\,\ref{fig:Fig1-JWT}.
Following the path, we define the following mapping between the spin operators and Majorana fermions
\begin{eqnarray}
 \label{eq:JWT}
 X_{j}=U_{j} \gamma^b_{j},\;\;\;\;
 Z_{j}=U_{j} \gamma^a_{j},\;\;\;\;
 Y_{j}=i \gamma^b_{j} \gamma^a_{j},
\end{eqnarray}
where $U_{j}=\prod_{j^{\prime}<j} Y_{j^{\prime}} $ is a non-local disorder operator attached to guarantee that Pauli operators at different sites commute. 
Here, $j$ labels the position of the spin along the Jordan-Wigner path. 
$j^{\prime}<j$ simply means the spin at $j^{\prime}$ is ordered before the spin at $j$ on the path.
$\gamma^a_{j}$ and $\gamma^b_{j}$ are Majorana fermion operators non-local in terms of the spin operators
\begin{eqnarray}
 \gamma^a_{j} &=& \big[ \prod_{j^{\prime}<j} Y_{j^{\prime}} \big] Z_{j},\label{eq:spin-MF1}\\
 \gamma^b_{j} &=& \big[ \prod_{j^{\prime}<j} Y_{j^{\prime}} \big] X_{j},\label{eq:spin-MF2}
 \label{eq:JWT-gamma}
\end{eqnarray}
and they satisfy the anti-commutation relationship
\begin{eqnarray}
 \{\gamma^a_{j^{\prime}},\gamma^a_{j}\}=\{\gamma^b_{j^{\prime}},\gamma^b_{j}\}=2\delta_{j^{\prime} j}, \;\{\gamma^a_{j^{\prime}},\gamma^b_{j}\}=0.
\end{eqnarray}

Now, we substitute Eq.\,(\ref{eq:JWT}) into the square plaquette operator defined in Eq.\,(\ref{eq:AK1}) and obtain
\begin{equation}
 \label{eq:AK1-JWT}
 A_k = \raisebox{-5.3mm}{ \includegraphics[scale=0.3]{Ak_plaquette.jpg} } = (U_1\gamma^b_1)(U_2 \gamma^a_2)(U_3\gamma^b_3)(U_4\gamma^a_4).
\end{equation}
Depending on the direction of the Jordan-Wigner path, there are two possible mappings of $A_k$. 
If the path goes from site $1$ to $4$, and $3$ to $2$ in a counter-clockwise fashion, then Eq.\,(\ref{eq:AK1-JWT}) becomes
\begin{equation}
\label{eq:AK1-JWT1}
 A_{k, \circlearrowleft} = (i\gamma^a_1\gamma^a_4)(i\gamma^a_3\gamma^a_2).
\end{equation}
Otherwise, we have
\begin{equation}
\label{eq:AK1-JWT2}
 A_{k, \circlearrowright} = (i\gamma^b_1\gamma^b_4)(i\gamma^b_3\gamma^b_2).
\end{equation}
Similarly, the pentagon plaquette operator defined in Eq.\,(\ref{eq:AK2}) can be rewritten as
\begin{equation}
\label{eq:AK2-JWT1}
 A^{\prime}_k= \raisebox{-5.5mm}{ \includegraphics[scale=0.3]{Ak2_plaquette.jpg} } = (U_1\gamma^b_1)(U_2 \gamma^a_2)(U_3\gamma^b_3)(U_4\gamma^a_4)(i\gamma^a_5\gamma^b_5),
\end{equation}
which reduces to
\begin{eqnarray}
 A^{\prime}_{k, \circlearrowleft}  = (i\gamma^a_1\gamma^a_4)(i\gamma^a_3\gamma^a_2), \label{eq:AK2-JWT21} \\
 A^{\prime}_{k, \circlearrowright} = (i\gamma^b_1\gamma^b_4)(i\gamma^b_3\gamma^b_2). \label{eq:AK2-JWT22}
 \end{eqnarray}
From Eqs.\,(\ref{eq:AK1-JWT1})-(\ref{eq:AK2-JWT22}), it is clear that each plaquette operator consists of the product of two pairwise link terms of a-type (b-type) Majoranas when the path ordering is counter-clockwise (clockwise).
For each site except the twist defects [site $5$ in Eq.\,(\ref{eq:AK2-JWT1})], both the a-type and b-type Majorana modes are involved in the plaquette operators because the top two plaquettes containing the site have an opposite path ordering of the bottom two plaquettes.
However, the twist defects are special because the Majorana modes associated with them do not show up in the pentagon plaquette operator.
Since the assignment of a-type ($\gamma^a$) and b-type ($\gamma^b$) Majorana fermions is arbitrary in Eq.\,(\ref{eq:JWT}), the two path orderings are equivalent upon exchanging $\gamma^a$ with $\gamma^b$.
Therefore, it is sufficient to assume the case of clockwise ordering in Eq.\,(\ref{eq:AK2-JWT22}).
In this case, $\gamma^a_5$ are involved in the two plaquettes below $A^{\prime}_k$ and $\gamma^b_5$ becomes an unpaired Majorana (it does not appear in the Hamiltonian.)
Clearly, introducing a twist defect gives rise to an unpaired Majorana mode in the new representation.

A pair of twists will produce two unpaired Majorana zero modes which can be used to encode qubit.
In Fig.\,\ref{fig:Fig1-JWT}, they correspond to $\gamma^b_{12}$ and $\gamma^b_{19}$
The parity operator associated with these two unpaired Majorana zero modes is given by
\begin{equation}
 P=i\gamma^b_{12}\gamma^b_{19}.
\end{equation}
Now, we want to express $P$ in the spin representation.
To eliminate the contribution from the boundary, we need to modify the form of the boundary spin operators in the Jordan-Wigner transformation.
Specifically, at sites $14$ and $15$ (Fig.\,\ref{fig:Fig1-JWT}), we exchange $X_{14}$ with $Y_{14}$, and $Z_{15}$ with $Y_{15}$ in Eqs.\,(\ref{eq:JWT})-(\ref{eq:JWT-gamma}). 
As a result, $X_{14}$ and $Z_{15}$ replace $Y_{14}$ and $Y_{15}$ in the definition of the disorder operators $U_{j>15}=(\prod_{i<14} Y_i) X_{14}Z_{15} (\prod_{j\ge k>15}Y_k)  $.

Using the modified Jordan-Wigner transformation from Eqs.\,(\ref{eq:spin-MF1})-(\ref{eq:spin-MF2}), the parity operator can be rewritten as
\begin{equation}
 P = i Z_{12}Y_{13}X_{14}Z_{15}Y_{16}Y_{17}Y_{18}X_{19}.
\end{equation}
We highlight the spin representation of $P$ in green ovals in Fig.\,\ref{fig:Fig1-JWT}.
Since each plaquette operator has an eigenvalue of $+1$ in the ground state, we can multiple $P$ by the plaquette operators shaded in dark brown in Fig.\,\ref{fig:Fig1-JWT}.
We end up with an equivalent parity operator shown in red circles in Fig.\,\ref{fig:Fig1-JWT}
\begin{equation}
  P^{\prime}=iX_{12}Y_{11}Y_{10}Z_9X_{20}Z_{19}.
  \label{eq:Parity-JWT}
\end{equation}
Compared to $P$, $P^{\prime}$ is a better representation because it does not go through the boundary and has a distance of roughly the separation of the two twist defects.
This is also the reason why we modify the Jordan-Wigner transformation; otherwise, we will have a term of $Z_{14}X_{15}$ from the boundary in the definition of $P^{\prime}$.
It is fairly easy to check that $P^{\prime}$ commutes with all the plaquette operators and it can not be written as the product of plaquette operators.
In fact, if we encode information in the degeneracy states due to the two twists, then $P^{\prime}$ is essentially the logical $Z$ operator accessing those states. 

\begin{figure}[tb]
\centering
\includegraphics[width=0.5\textwidth]{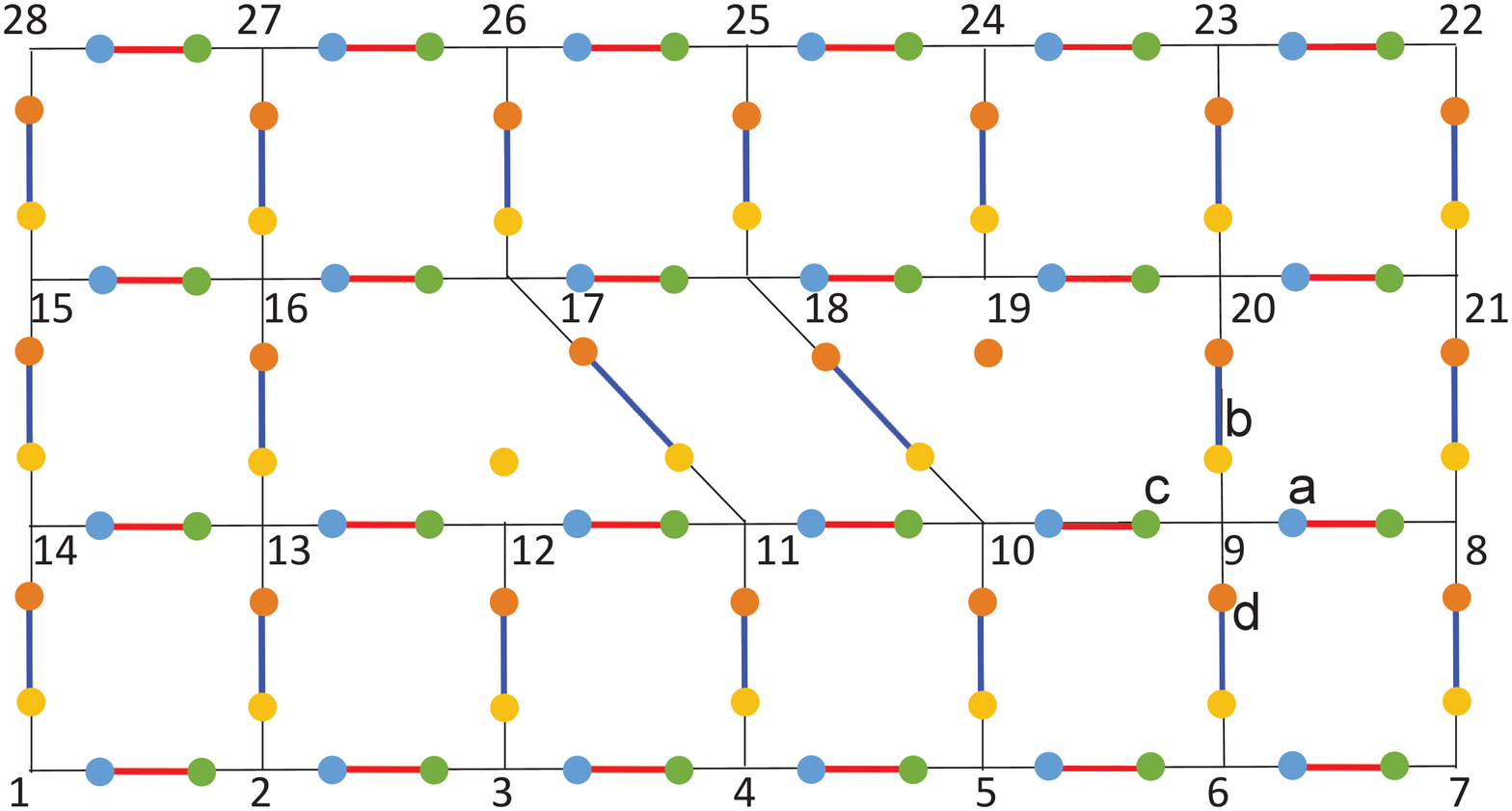}
\caption{(color online) A Majorana fermion model on a 2D lattice with twist defects. Each site has four Majorana modes (dots): $a$, $b$, $c$, and $d$. 
$a$ and $c$ modes form bonds (red) in the horizontal direction, while $b$ and $d$ modes form bonds (green) in the vertical direction}
\label{fig:Fig2-MFModel}
\end{figure}

\begin{figure}[h]
\centering
\includegraphics[width=0.47\textwidth]{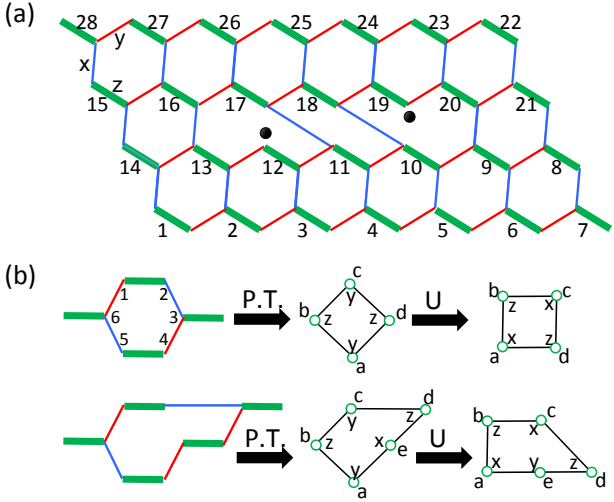}
\caption{(color online) Perturbative limit of honeycomb lattice model with twist defects. (a) The Kitaev honeycomb model with two twist dislocations. The unpaired Majorana modes are highlighted by the dots. 
(b) In the strong z-bond limit (think green bonds), under perturbation theory (P.T.) the hexagonal (octagon) plaquette maps to the square (pentagon) plaquette. 
A local unitary transformation $U$ is needed to bring the plaquette operators into the canonical form.
}
\label{fig:Fig3-HCModel}
\end{figure}

\subsection{Projective construction from a Majorana surface code model}
\label{sec:MFModel}

Inspired by the projective construction due to \citet{WenPRL03}, here we start with a surface code model of Majorana fermions, show there are unpaired Majorana modes at the twist defects, and then map it to the surface code model of spins via projection.
The model is defined on a 2D planar square lattice with four Majorana modes $\gamma^{a,b,c,d}_n$ placed on each site $n$.
Each site forms a pairwise bond of Majorana modes with its nearest neighbors in the form shown in Fig.\,\ref{fig:Fig2-MFModel}.
Basically, $\gamma^a$ and $\gamma^c$ are paired along horizontal bonds, and $\gamma^b$ and $\gamma^d$ are paired along vertical bonds.
By construction, the Hamiltonian takes the form
\begin{equation}
 H_{\text{MF}}=-\sum_k A_k,
\end{equation}
where $A_k$ is plaquette operator of Majorana modes. For a square plaquette, it reads
\begin{equation}
\label{eq:AK1-MFM}
 A_k = \raisebox{-5.3mm}{ \includegraphics[scale=0.3]{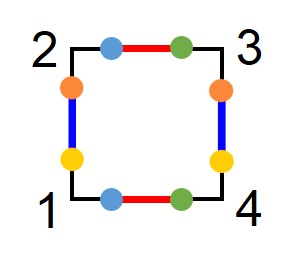} } = (i\gamma^b_1\gamma^d_2)(i\gamma^a_2\gamma^c_3)(i\gamma^d_3\gamma^b_4)(i\gamma^c_4\gamma^a_1).
\end{equation}
For a pentagon plaquette with a twist defect, it reads
\begin{equation}
\label{eq:AK2-MFM}
 A^{\prime}_k = \raisebox{-5.3mm}{ \includegraphics[scale=0.3]{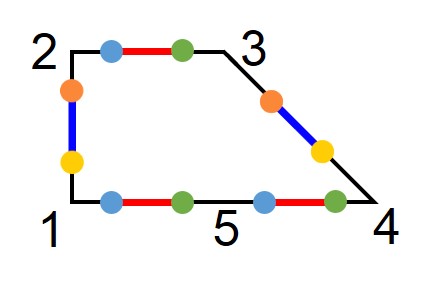} } = (i\gamma^b_1\gamma^d_2)(i\gamma^a_2\gamma^c_3)(i\gamma^d_3\gamma^b_4)(i\gamma^c_4\gamma^a_5)(i\gamma^c_5\gamma^a_1).
\end{equation}
This model is exactly solvable because each link operator $u^{ac}_{mn}=i\gamma^a_m\gamma^c_n$ or $u^{bd}_{mn}=i\gamma^b_m\gamma^d_n$ commutes with the Hamiltonian.
Since each link operator has two eigenvalues of $\pm1$, the ground state is the common eigenstate of all $A_k$ operators with eigenvalue $+1$.

Note that in Eq.\,(\ref{eq:AK2-MFM}), $\gamma^b_5$ is absent from the pentagon plaquette operator and it is not involved in the two square plaquettes below $A^{\prime}_k$ either. 
Therefore, $\gamma^5_b$ is an unpaired Majorana mode associated with the twist defect.
To perform the mapping to spins, we first define two fermion operators out of the four Majorana modes at each site
\begin{equation}
 \psi_{\alpha n}=\frac{\gamma^a_n+i\gamma^d_n}{2},\;\;\psi_{\beta n}=\frac{\gamma^c_n+i\gamma^b_n}{2}.
\end{equation}
The Hilbert space at site $n$ is spanned by the occupation of the two fermion modes: $|00\rangle_n$, $|01\rangle_n$, $|10\rangle_n$, and $|11\rangle_n$.
The fermion parity of site $n$ is given by
\begin{equation}
 D_n = (2\psi^{\dagger}_{\alpha n}\psi_{\alpha n}-1)(2\psi^{\dagger}_{\beta n}\psi_{\beta n}-1)=\gamma^a_n\gamma^b_n\gamma^c_n\gamma^d_n.
\end{equation}
The surface code model of spins can be obtained by projecting to the subspace of even parity: $D_n=1$. 
This reduces the dimension of each site from four to two.
We can map the allowed states $\{|00\rangle_n, |11\rangle_n\}$ to effective spin states: $\mid \Uparrow\rangle_n\equiv |00\rangle_n$ and $\mid \Downarrow\rangle_n\equiv |11\rangle_n$.
Under this projection, the Hamiltonian described by Eqs.\,(\ref{eq:AK1-MFM})-(\ref{eq:AK2-MFM}) can be rewritten in terms of the effective spin Pauli operators.
For instance, $i\gamma^a_1\gamma^b_1|00\rangle_1=|11\rangle_1/4$ and $i\gamma^a_1\gamma^b_1|11\rangle_1=|00\rangle_1/4$, therefore $i\gamma^a_1\gamma^b_1$ maps to $S^x_1/4$.
It is straightforward to carry out this mapping and show that the surface code model of Majorana fermions indeed reduces to the spin model in Eqs.\,(\ref{eq:AK1})-(\ref{eq:AK2}) after projecting to the even-parity subspace.

For the two unpaired Majorana modes shown in Fig.\,\ref{fig:Fig2-MFModel}, the total fermion parity is given by $P=i\gamma^b_{12}\gamma^d_{19}$.
However, after the projection, $P$ is no longer a physical operator because it does not commute with the fermion parity operator $D_{12}$ and $D_{19}$, namely it brings the system out of the even-parity subspace.
A simple cure to this problem is attaching a string operator $S_{12,19}$ connecting two unpaired Majorana modes.
One possible choice of such a string operator is the product of all the link operators going from site $12$ to site $19$
\begin{equation}
 P^{\prime} = i \gamma^b_{12}\gamma^d_{19} S_{12,19}, \;\; S_{12,19}=u^{ac}_{12,11} u^{ac}_{11,10} u^{ac}_{10,9} u^{bd}_{9,20} u^{ac}_{20,19}.
\label{eq:Parity-MFModel}
 \end{equation}
This newly defined parity operator communtes with the fermion parity operator $D_n$. 
Using the mapping from Majorana operators to Pauli operators under even-parity constraint, Eq.\,(\ref{eq:Parity-MFModel}) reduces to the expression of Eq.\,(\ref{eq:Parity-JWT}) obtained using Jordan-Wigner transformation.

\subsection{Perturbative limit of the Kitaev honeycomb model}
The surface code model can be obtained from the perturbative limit of the Kitaev honeycomb model.
It has been shown that the Kitaev honeycomb model admits unpaired Majorana modes after introducing dislocations into the lattice \cite{PetrovaPRB13,PetrovaPRB14}.
Naturally, one would expect the unpaired Majorana zero modes associated with twist defects in the toric code model can be deduced from the Kitaev honeycomb model with a certain type of dislocations.
Indeed, we show this is the case by starting from the honeycomb model with dislocations and then performing the perturbation theory.

The Kitaev honeycomb model has spin-$1/2$ particles interacting through two-body terms \cite{KitaevAP06,PetrovaPRB14}
\begin{equation}
 H_{\text{HCM}}=-J_x\sum_{x \text{links}}\sigma^x_m \sigma^x_n-J_y\sum_{y \text{links}}\sigma^y_m \sigma^y_n-J_z\sum_{z \text{links}}\sigma^z_m \sigma^z_n,
\end{equation}
where the classification of $x,y,z$ links is shown in Fig.\,\ref{fig:Fig3-HCModel}(a).
This model can be exactly solved by mapping spin $\sigma^{x,y,z}_n$ to four Majorana modes $\gamma^{x,y,z}_n$ and $\eta_n$
\begin{equation}
\label{eq:HCM-spin-MF}
 \sigma^{\alpha}_n = i\gamma^{\alpha}_n \eta_n,
\end{equation}
with the Majorana modes satisfying the anti-commutation relation
\begin{equation}
 \{\gamma^{\alpha}_m,\gamma^{\beta}_n\}=2\delta_{\alpha\beta}\delta_{mn},\{\eta_m,\eta_n\}=2\delta_{mn},\{\gamma_m^{\alpha}, \eta_n\}=0.
\end{equation}
The dimension of the Hilbert space at each site is increased from two to four after the mapping.
Such a problem is fixed by defining a projection operator
\begin{equation}
 D_n\equiv \gamma^x_n\gamma^y_n\gamma^z_n \eta_n,
\end{equation}
and constraining the physical states to be eigenstates of $D_n$ with eigenvalue $+1$.
The Hamiltonian can be rewritten in terms of Majorana modes
\begin{equation}
 H_{\text{HCM}}=i\sum_{\alpha=x,y,z}\sum_{\langle mn \rangle}J_{\alpha}u^{\alpha}_{mn}\eta_m\eta_n,
\end{equation}
where $\langle mn\rangle$ denotes the bond connecting spins $m$ and $n$ and the link operator $u^{\alpha}_{mn}=i\gamma^{\alpha}_{m}\gamma^{\alpha}_{n}$.
It is straightforward to check that $u^{\alpha}_{mn}$ commutes with the Hamiltonian and hence can be set to either $+1$ or $-1$.
The choice of the value of the link operator fixes the $Z_2$ background gauge.
As a result, the Hamiltonian becomes quadratic in terms of $\eta$ Majorana modes and can be solved exactly \cite{KitaevAP03,PetrovaPRB14}.

A pair of twist dislocations can be added by modifying the bonds according to Fig.\,\ref{fig:Fig3-HCModel}(a).
Every $\gamma$ Majorana mode is paired with another $\gamma$ in the link operator except $\gamma^x_{12}$ and $\gamma^x_{19}$, both of which disappear from the Hamiltonian and hence are unpaired Majorana modes.
Together they form a fermion mode introducing an additional topological degeneracy of two.
Next, we show that in the strong $z-$bond limit the honeycomb model with dislocations maps to the surface code model with twist defects, and the unpaired Majorana modes carry over.

In the limit $J_z\gg J_x,J_y>0$, it becomes energetically favorable to have the spins in the $z-$bonds aligned, i.e., $\sigma^z_m\sigma^z_n=1$.
Each $z-$bond becomes an effective spin with states $\mid\Uparrow\rangle \equiv \mid\uparrow\uparrow\rangle_{mn}$ and $\mid\Downarrow\rangle \equiv \mid\downarrow\downarrow\rangle_{mn}$.
In this case, the Hamiltonian can be split into two parts
\begin{eqnarray}
 H_{\text{HCM}}&=&H_0+V, \\
 H_0&=&-J_z\sum_{z \text{links}}\sigma^z_m \sigma^z_n, \\
 V&=&-J_x\sum_{x \text{links}}\sigma^x_m \sigma^x_n-J_y\sum_{y \text{links}}\sigma^y_m \sigma^y_n.
\end{eqnarray}
$H_0$ restricts the ground state to the subspace of $\mid\Uparrow\rangle $ and $\mid\Downarrow\rangle $ and $V$ can be treated as a perturbation.
For a hexagonal plaquette without dislocations as shown in Fig.\,\ref{fig:Fig3-HCModel}(b), the first non-vanishing term from the $4^{\text{th}}$-order perturbation theory \cite{KitaevAP03} is
\begin{eqnarray}
 H^{(4)}_{\text{eff}} &=& \Upsilon^{\dagger} V G_0 V G_0 V G_0 V \Upsilon = -\frac{J^2_xJ^2_y}{16 J^3_z}\sum_p Q^{(4)}_p+\text{const.}, \nonumber \\
 Q^{(4)}_p &=& (\sigma^y_6\sigma^y_1)(\sigma^x_2\sigma^x_3)(\sigma^y_3\sigma^y_4)(\sigma^x_5\sigma^x_6),
 \end{eqnarray}
where $G_0=(E_0-H_0)^{-1}$ is the non-interacting Green function with $E_0$ being the ground state energy of $H_0$, and $\Upsilon$ is the projection operator of the subspace $\{\mid\Uparrow\rangle,\mid\Downarrow\rangle\}$.
Regrouping the Pauli operators in $Q^{(4)}_p$ enables us to rewrite $Q^{(4)}_p$ in the effective spin representation
\begin{equation}
 Q^{(4)}_p=-(\sigma^y_4\sigma^x_5)(\sigma^x_6\sigma^y_6)(\sigma^y_1\sigma^x_2)(\sigma^x_3\sigma^y_3)\equiv S^y_a S^z_b S^y_c S^z_d,
\end{equation}
where $S^{y,z}$ is the effective Pauli operator for the strong z-bonds shown in Fig.\,\ref{fig:Fig3-HCModel}(b).
For a plaquette with a twist dislocation as shown in Fig.\,\ref{fig:Fig3-HCModel}(b), we follow the same procedure and the $5^{\text{th}}$-order perturbation theory gives rise to the following Hamiltonian in the effective spin representation
\begin{eqnarray}
 H^{(5)}_{\text{eff}} &=& -\frac{5J^2_xJ^3_y}{128J^4_z}\sum_p Q^{(5)}_p+\text{const.}, \nonumber \\
 Q^{(5)}_p &=& S^y_a S^z_b S^y_c S^z_d S^x_e.
\end{eqnarray}
A local unitary transformation $U=\prod_m e^{-i\frac{\pi}{4}S^z_m}$ will bring $Q^{(4)}_p$ and $Q^{(5)}_p$ into the same form as the surface code model introduced in Eqs.\,(\ref{eq:AK1})-(\ref{eq:AK2}).
Therefore, after the perturbation theory and the transformation $U$, the honeycomb model in Fig.\,\ref{fig:Fig3-HCModel}(a) reduces to the surface code model in Fig.\,\ref{fig:Fig1-JWT}.
The unpaired Majorana modes $\gamma^x_{12}$ and $\gamma^x_{19}$ associated with twist dislocations in the honeycomb model carry over to the surface code model.
Similar to the case of Majorana fermion model in Sec.\,\ref{sec:MFModel}, $i\gamma^x_{12}\gamma^x_{19}$ is not the physical parity operator because it does not commute with the projection operator $D_{12}$ and $D_{19}$.
Again, the solution is to attach the link operator $u^{\alpha}_{mn}$ connecting site $12$ with site $19$ to $i\gamma^x_{12}\gamma^x_{19}$ so that it commutes with $D_n$.
In the perturbative limit, this newly defined parity operator reduces to the parity operator we found previously in surface code mode. It is highlighted in red circles in Fig.\,\ref{fig:Fig1-JWT}.

\begin{figure}[tb]
\centering
\includegraphics[width=0.5\textwidth]{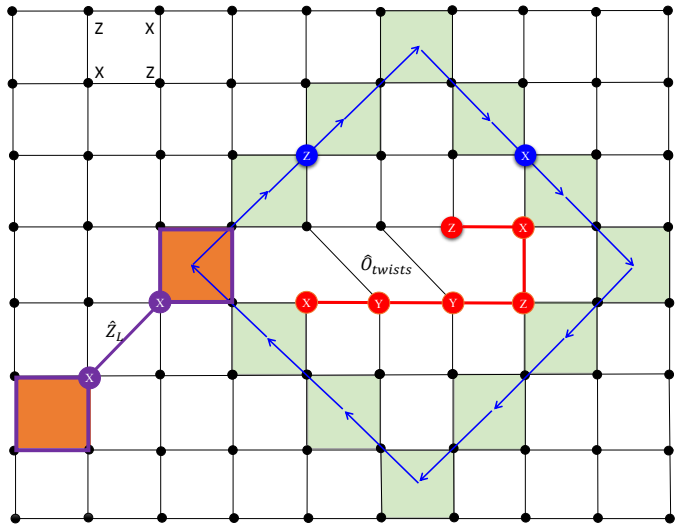}
\caption{(color online) Measurement of the parity (topological charge) of two twist defects in the surface code stabilizer setting. The direct way is to measure simultaneously the Pauli operators (red circles) making up the parity operator.
Another indirect approach is to create a measurement qubit of two holes by stopping two stabilizers shaded in dark brown.
Moving one of the hole around the twists encodes the parity in the logical $Z_L$ of the measurement qubit (product of the X's connecting the two holes).
By measuring the logical $Z_L$ before and after moving the hole, the parity of twists can be read out.}
\label{fig:Fig4-ParityM}
\end{figure}

\section{Non-Abelian Statistics of Majorana Fermions}
\label{Non-Abelian}
In this section, we focus on demonstrating the non-Abelian
statistics of Majorana fermions using the twist defects through braiding operations. 
Since introducing the twists requires surgery of the underlying lattice Hamiltonian, moving twists would constantly change the Hamiltonian, which is not favorable from the point of view of experiments.
Instead, we propose to use measurement-based braiding \cite{BondersonPRL08,BondersonPRB13} to show the non-Abelian statistics.
Our proposal requires only the parity measurements of a pair of twists.
We first demonstrate how to perform such parity measurements in a surface code setting.
Then we elaborate on the details of measurement-based braiding, and show that a single cycle of parity measurements followed by a single logical qubit operation is sufficient to simulate the braiding operation.
This removes the uncertainty of number of measurements in the forced measurement scheme \cite{BondersonPRL08}.

\subsection{Parity measurement of twist defects}
\label{sec:Parity_Measure}
All of our previous discussion is based on the Hamiltonian formulation, and it is still very challenging to engineer the four-body or five-body plaquette operators in Eqs.\,(\ref{eq:AK1})-(\ref{eq:AK2}).
Alternatively, one can use the stabilizer formulation in topological codes \cite{BravyiArXiv98,FreedmanFCM01,DennisJMP02,FowlerPRA12}.
In this case, there is no Hamiltonian and each term of the Hamiltonian is replaced by a stabilizer operator. 
The stabilizers can be implemented using a combination of single-qubit gates and two-qubit CNOT gates followed by a projective measurement; see \citet{FowlerPRA12} for details.
In each cycle of the code, all the stabilizers are measured simultaneously.
The resulting state after a projective measurement is an eigenstate of all stabilizers with randomly selected eigenvalue of either $+1$ or $-1$.
Error detection is done through repeating the cycle of stabilizer measurements, comparing the measurement results, and identifying specific errors on particular qubits with the help of classical matching algorithms.

To demonstrate the non-Abelian statistics of twists, we need to perform parity measurement frequently.
We propose to do such parity measurement in a stabilizer setting as shown in Fig.\,\ref{fig:Fig4-ParityM}.
The first method would be directly measuring the parity operator $\hat{O}_{\text{twists}}$ that is a string of Pauli operators highlighted in red circles in Fig.\,\ref{fig:Fig4-ParityM}.
This can be done in three steps. First, we turn off the stabilizers involving those Pauli operators to isolate them.
Second, we perform corresponding Pauli $X/Y/Z$ measurement on individual qubits.
The product of the measurement outcomes is the parity.
Finally, we turn the stabilizers back on. Error detection is done by comparing the stabilizer measurements before and after turning off the stabilizers.

A second indirect approach to measure the parity is to create an ancillary measurement logical qubit by stopping measuring two stabilizers shaded in dark brown in Fig.\,\ref{fig:Fig4-ParityM}.
This creates two holes, and generates a four-dimensional degree of freedom with each hole taking on an eigenvalue of $+1$ or $-1$.
We effectively encode a single qubit in this four-dimensional subspace by defining the logical $\hat{Z}_L$ as the operator connecting the two holes and logical $\hat{X}_L$ as the stabilizer of one of the two holes. 
Then, we move one of the holes (the top right one in Fig.\,\ref{fig:Fig4-ParityM}) around the two twists forming a loop.
This process performs a CNOT gate between the ancillary qubit and the qubit encoded in two twists \cite{FowlerPRA12,VijayArXiv15}
\begin{equation}
 \hat{Z}_L \rightarrow \hat{Z}_L \otimes \hat{O}_{\text{twists}}.
\end{equation}
Therefore, by comparing the outcome of $\hat{Z}_L$ measurement before and after moving the hole around the twists, we can read out the qubit encoded in the two twists.
Note that such a measurement is quantum non-demolitional and can be repeated several times to improve the accuracy of the parity readout.

\begin{figure}[tb]
\centering
\includegraphics[width=0.49\textwidth]{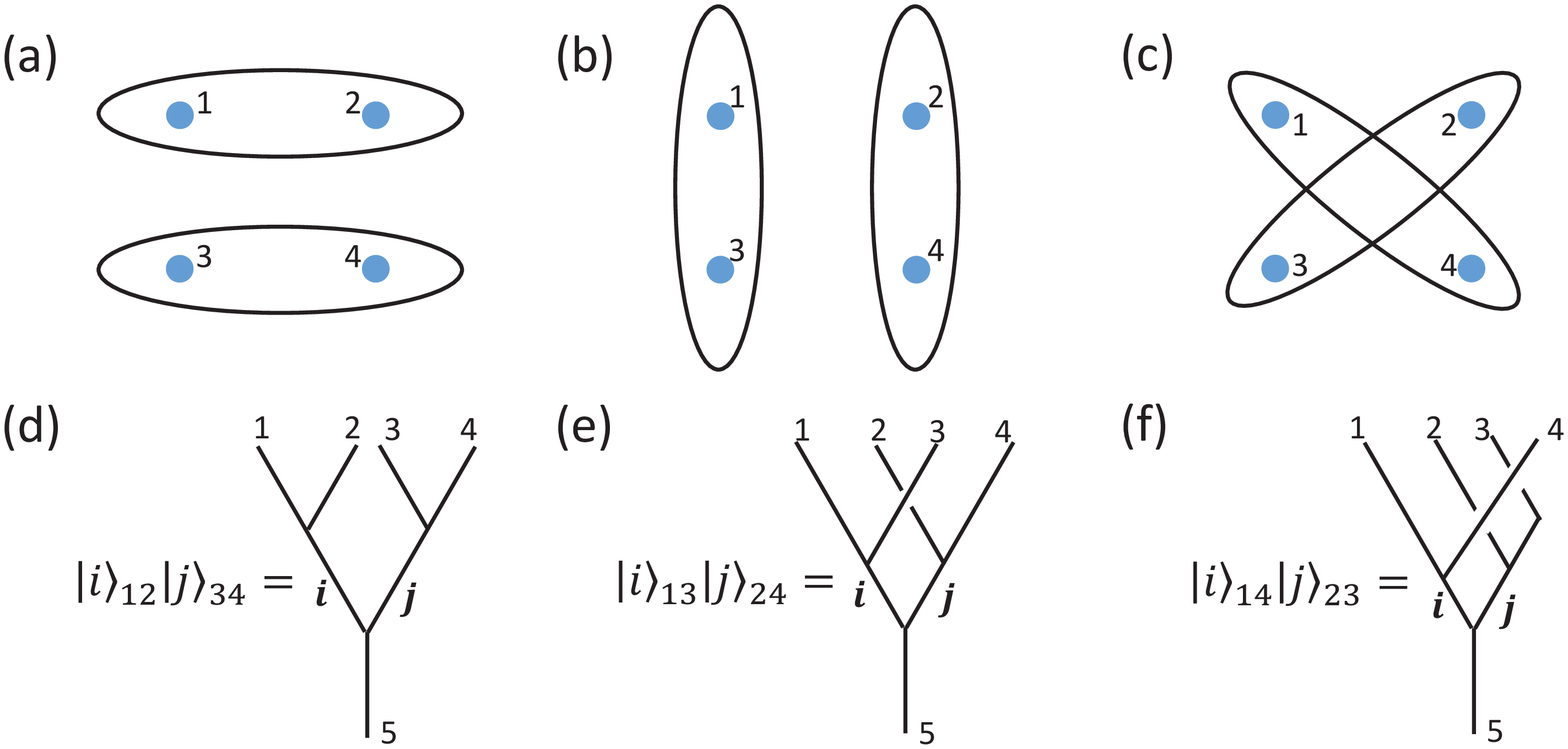}
\caption{(color online) Three different bases of four Majorana modes. (a)-(c) Three different ways to group Majoranas $1$ to $4$. (d)-(f) The corresponding fermion number states expressed in fusion tree diagrams.}
\label{fig:Fig5-basis}
\end{figure}

\subsection{Basis transformation of Majorana fermions}

Now, we work out the details of measurement-based braiding by following the wave-function after each projective measurement in the Schr\"{o}dinger picture.
Because the scheme involves measuring the topological charge of anyons in different bases, a basis transformation between different ways of grouping Majorana modes is needed.
Given a set of four Majorana zero modes, we have three distinct ways to group them into two pairs with fermion charge $i$ and $j$ as shown in Fig.\,\ref{fig:Fig5-basis}(a)-(c). 
These basis states can be represented using fusion tree diagrams shown in Fig.\,\ref{fig:Fig5-basis}(d)-(f). 
The Majorana fermions can be treated as Ising anyons (up to an overall irrelevant phase factor in the braiding operator). 
The Ising anyon model has three types of topological charges: vacuum $I$, Ising anyon $\sigma$ and fermion $\psi$.
The fusion rule of Ising anyon model \cite{Pachos12} is given by
\begin{eqnarray}
 I\times I&=&I,\;I\times\sigma=\sigma,\;I\times\psi=\psi, \nonumber \\
 \sigma \times \sigma &=& I +\psi,\; \sigma\times \psi=\sigma,\; \psi\times \psi=I.
\end{eqnarray}
Depending on the total fermion parity of the four Majoranas, the charges $i$ and $j$ further fuse into anyon $5$ which is either vacuum $I$ for even parity or a fermion $\psi$ for odd parity \cite{Pachos12}.
The basis transformation can be viewed as changing the ordering of fusion.
This amounts to transform the fusion tree diagrams in Fig.\,\ref{fig:Fig5-basis}(d)-(f) into each other.
Such a transformation is dictated by the $F$-matrices, $R$-matrices and $B$-matrices as following
\begin{eqnarray}
 \raisebox{-9mm}{ \includegraphics[scale=0.3]{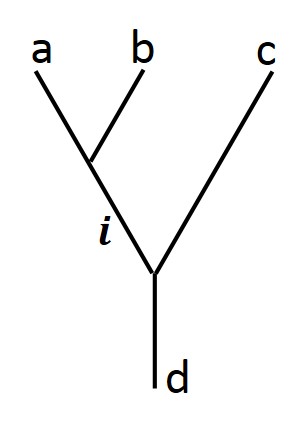} } &=& \sum_j[F^{d}_{abc}]_{ij} \raisebox{-9mm}{ \includegraphics[scale=0.3]{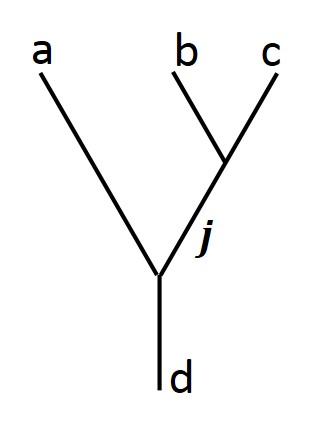} },\nonumber \\
 \raisebox{-9mm}{ \includegraphics[scale=0.3]{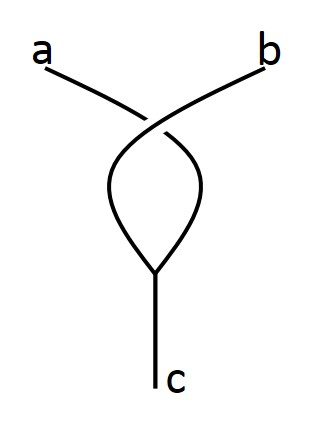} } &=& R^c_{ab} \raisebox{-9mm}{ \includegraphics[scale=0.3]{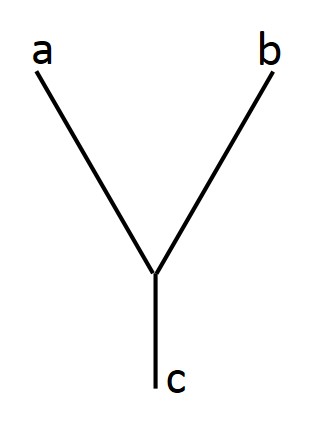} },\nonumber \\
 \raisebox{-9mm}{ \includegraphics[scale=0.3]{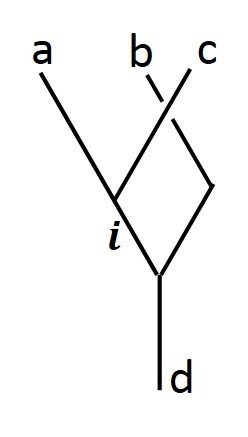} } &=& \sum_j[B^{d}_{abc}]_{ij} \raisebox{-9mm}{ \includegraphics[scale=0.3]{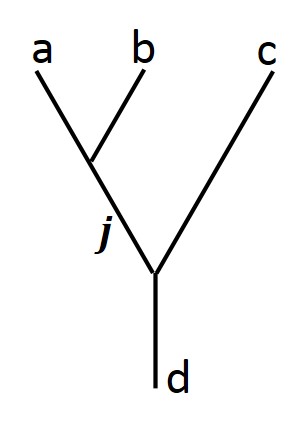} },
\label{eq:FBTree}
 \end{eqnarray}
where the $F$, $R$ and $B$ ($=F^{-1}RF$) matrices for Ising anyons are given by \cite{Pachos12}
\begin{eqnarray}
 &&  
 [F^{\sigma}_{\sigma\sigma\sigma}] = \frac{1}{\sqrt{2}} {\begin{bmatrix} 1 & 1 \\ 1 & -1 \end{bmatrix}},  \nonumber \\
 && [R_{\sigma\sigma}]= e^{-i\frac{\pi}{8}}\begin{bmatrix} 1 & 0 \\ 0 & i \end{bmatrix}, 
[B^{\sigma}_{\sigma\sigma\sigma}] = \frac{e^{i\frac{\pi}{8}}}{\sqrt{2}}\begin{bmatrix} 1 & -i \\ -i & 1 \end{bmatrix}.
 \label{eq:FRB-matrices}
 \end{eqnarray}
Here, the matrices are all in the basis of $\{I,\psi\}$. 
$[F^{\psi}_{\sigma\psi\sigma}]_{\sigma\sigma}=[F^{\sigma}_{\psi\sigma\psi}]_{\sigma\sigma}=-1$. 
All the other matrix elements are either $1$ if it is allowed by the fusion rules or $0$ if not allowed.

We first work with the even-parity subspace which means anyon $5$ after all the fusions is vacuum $I$ in Fig.\,\ref{fig:Fig5-basis}(d)-(f).
As an example, the steps to transform from the basis $(13,24)$ to $(12,34)$ are given by
\begin{equation}
\centering
\includegraphics[width=0.49\textwidth]{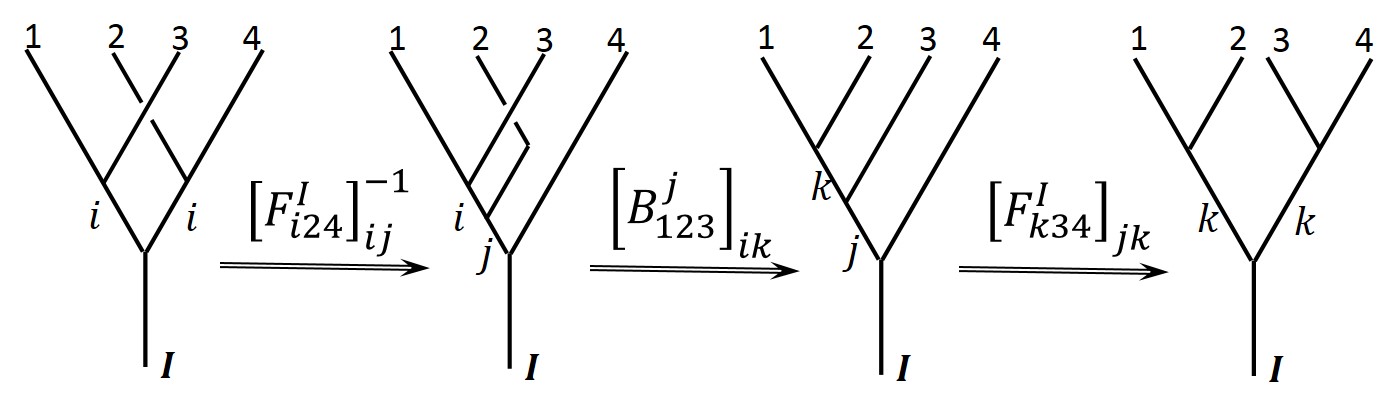}. \nonumber
\end{equation}
Mathematically, this transformation can be expressed as
\begin{equation}
 |i\rangle_{13}|i\rangle_{24} = \sum_{jk} [F^I_{i24}]^{-1}_{ij}[B^j_{123}]_{ik}[F^I_{k34}]_{jk} |k\rangle_{12}|k\rangle_{34},
\end{equation}
where $|k\rangle_{ab}$ denotes the wave-function of the state after fusing anyons $a$ and $b$ into a channel with topological charge $k$.
Here, $1-4$ are Ising anyons $\sigma$, $j$ is $\sigma$ according to the fusion rules, and $k$ can be either $I$ or $\psi$.
Plugging in the matrices from Eq.\,(\ref{eq:FRB-matrices}) gives the following result
\begin{eqnarray}
&& \begin{bmatrix} |0\rangle_{13}|0\rangle_{24} \\ |1\rangle_{13}|1\rangle_{24} \end{bmatrix} = U^{(e)}_{13\leftarrow 12} \begin{bmatrix} |0\rangle_{12}|0\rangle_{34} \\ |1\rangle_{12}|1\rangle_{34} \end{bmatrix}, \nonumber \\
&& U^{(e)}_{13\leftarrow 12} = \frac{e^{i\frac{\pi}{8}}}{\sqrt{2}}\begin{bmatrix} 1 & -i \\ -i & 1 \end{bmatrix}.
\label{eq:Ue}
\end{eqnarray}
Repeat the above procedure, we obtain the unitary transformations between $(14,23)$ and $(12,34)$, and between $(13,24)$ and $(14,23)$ in the even-parity subspace
\begin{eqnarray}
 && U^{(e)}_{14\leftarrow 12} = \frac{1}{\sqrt{2}}\begin{bmatrix} 1 & 1 \\ -i & i \end{bmatrix}, \nonumber \\
 && U^{(e)}_{13\leftarrow 14} = \frac{e^{-i\frac{\pi}{8}}}{\sqrt{2}}\begin{bmatrix} 1 & -1 \\ 1 & 1 \end{bmatrix}.
\end{eqnarray}

Similarly, the basis transformation in the odd-parity subspace is obtained in the same way as
\begin{eqnarray}
 && \begin{bmatrix} |0\rangle_{13}|1\rangle_{24} \\ |1\rangle_{13}|0\rangle_{24} \end{bmatrix} = U^{(o)}_{13\leftarrow 12} \begin{bmatrix} |0\rangle_{12}|1\rangle_{34} \\ |1\rangle_{12}|0\rangle_{34} \end{bmatrix}, \nonumber \\
&& U^{(o)}_{13\leftarrow 12} = \frac{e^{i\frac{\pi}{8}}}{\sqrt{2}}\begin{bmatrix} 1 & -i \\ -i & 1 \end{bmatrix}, \nonumber \\
&& U^{(o)}_{14\leftarrow 12} = \frac{1}{\sqrt{2}}\begin{bmatrix} i & -i \\ 1 & 1 \end{bmatrix}, \nonumber \\
 && U^{(o)}_{13\leftarrow 14} = \frac{e^{-i\frac{\pi}{8}}}{\sqrt{2}}\begin{bmatrix} 1 & 1 \\ -1 & 1 \end{bmatrix}.
\label{eq:Uo}
 \end{eqnarray}
It is straightforward to check that the unitary transformations satisfy the consistency equation
\begin{eqnarray}
&& U^{(e)}_{13\leftarrow 12} = U^{(e)}_{13\leftarrow 14} U^{(e)}_{14\leftarrow 12}, \nonumber \\
&& U^{(o)}_{13\leftarrow 12} = U^{(o)}_{13\leftarrow 14} U^{(o)}_{14\leftarrow 12}.
\end{eqnarray}

\begin{figure}[tb]
\centering
\includegraphics[width=0.35\textwidth]{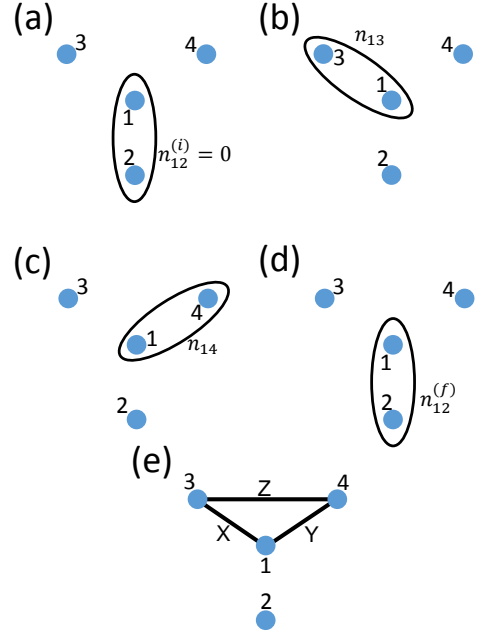}
\caption{(color online) Measurement-based braiding involving four Majorana modes. (a) The initialization of Majorana $1$ and $2$ in the vacuum state. (b)-(d) Measurements of the fermion number of Majorana $1$ and $3$, $1$ and $4$, and $1$ and $2$, respectively.
(e) Definition of the single-qubit logical $X$, $Y$ and $Z$ operators.}
\label{fig:Fig6-TCM}
\end{figure}

\subsection{Measurement-based braiding}
With the unitary transformations between bases in both the even-parity and odd-parity subspaces, we are ready to work out the details of measurement-based braiding (MBB) in the Schr\"odinger picture.
The measure-only approach to topological quantum computation was proposed by \citet{BondersonPRL08} to implement the topological gates without physically braiding the computational anyons.
The braiding operation instead is replaced by a series of quantum non-demolitional
topological charge measurements as shown in Fig.\,\ref{fig:Fig6-TCM}(a)-(d).

We consider the special case of Ising anyons (Majorana fermions) only. 
In our case, the topological charge (fermion-parity) of each pair of Majorana modes can be measured using the scheme proposed in Sec.\,\ref{sec:Parity_Measure}.
Anyons $3$ and $4$ are the computational anyons to be exchanged, and $1$ and $2$ are auxiliary anyons employed to assist MBB. 
The scheme works in the following way: one first initialize the two anyons $1$ and $2$ in the vacuum state with $n^{(i)}_{12}=0$ as shown in Fig.\,\ref{fig:Fig6-TCM}(a). 
Then a forced measurement is perform on anyons $1$ and $3$: if $n_{13}$ is not $0$, we go back to measure $n_{12}$ followed by $n_{13}$ and repeat until we obtain $n_{13}=0$ [Fig.\,\ref{fig:Fig6-TCM}(b)].
In some sense, we force anyons $1$ and $3$ to fuse into the vacuum sector by repeated measurements.
Physically, this procedure realizes anyonic teleportation of the state encoded in anyon $3$ to anyon $2$.
Similarly, forced measurements are done on anyons $1$ and $4$, and $1$ and $2$ shown in Fig.\,\ref{fig:Fig6-TCM}(c)-(d) to have $n_{14}=0$ and $n^{(f)}_{12}=0$, respectively.
It turns out the resulting effect is equivalent to a braiding operation on anyons $3$ and $4$ \cite{BondersonPRL08}.
Due to the inherent probabilistic nature of forced measurements, the operation time of each measurement-based braiding unavoidably varies from run to run.
This imposes difficulties of synchronizing the clock if one wants to use MBB to perform topological gates.
Recently, the generic case of all possible intermediate measurement results is considered \cite{BondersonPRB13}.
Here, it is our aim to study the generic case in detail and we find that one can perform MBB with a fixed number of measurements removing the uncertainty associated with forced measurements.

We assume that after the initialization step in Fig.\,\ref{fig:Fig6-TCM}(a), the state of the four anyons is described by
\begin{equation}
 |\Psi^{(a)}\rangle = |0\rangle_{12}\big[\alpha|0\rangle_{34} + \beta|1\rangle_{34} \big].
\end{equation}
Using the unitary transformations between bases in Eqs.\,(\ref{eq:Ue})-(\ref{eq:Uo}), we can rewritten the initial state in the basis of $(13,24)$
\begin{eqnarray}
 && |\Psi^{(a)}\rangle = \frac{\text{e}^{-i\pi/8}}{\sqrt{2}} \big[ \alpha(|0\rangle_{13}|0\rangle_{24}+i|1\rangle_{13}|1\rangle_{24})  \nonumber \\
 && \qquad \qquad \qquad \quad   + \beta(|0\rangle_{13}|1\rangle_{24}+i|1\rangle_{13}|0\rangle_{24}) \big].
\end{eqnarray}
After the projective measurement of $n_{13}$ in Fig.\,\ref{fig:Fig6-TCM}(b), depending on the outcome the state is
\begin{equation}
 |\Psi^{(b)}\rangle = 
    \begin{cases}
     \text{e}^{-i\frac{\pi}{8}}|0\rangle_{13} \big[\alpha|0\rangle_{24} + \beta|1\rangle_{24} \big], & \text{if } n_{13}=0 \\
    i\text{e}^{-i\frac{\pi}{8}}|1\rangle_{13} \big[\alpha|1\rangle_{24} + \beta|0\rangle_{24} \big], & \text{if } n_{13}=1 
    \end{cases}
\end{equation}
Next, we rewrite $|\Psi^{(b)}\rangle$ in the basis of $(14,23)$ and perform projective measurement of $n_{14}$ shown in Fig.\,\ref{fig:Fig6-TCM}(c).
The resulting state is
\begin{equation}
 |\Psi^{(c)}\rangle = 
 \begin{cases}
     \text{e}^{-i\frac{\pi}{4}}|0\rangle_{14} \big[\alpha|0\rangle_{23} + \beta|1\rangle_{23} \big], \quad n_{13}=n_{14}=0 \\
     \text{e}^{-i\frac{\pi}{4}}|1\rangle_{14} \big[-\alpha|1\rangle_{23} + \beta|0\rangle_{23} \big], n_{13}=\bar{n}_{14}=0 \\
    i\text{e}^{-i\frac{\pi}{4}}|0\rangle_{14} \big[\alpha|0\rangle_{23} - \beta|1\rangle_{23} \big], \quad n_{13}=\bar{n}_{14}=1 \\
    i\text{e}^{-i\frac{\pi}{4}}|1\rangle_{14} \big[\alpha|1\rangle_{23} + \beta|0\rangle_{23} \big], \quad n_{13}=n_{14}=1 
    \end{cases}
\end{equation}
where $\bar{n}=1-n$. Finally, we transform $|\Psi^{(c)}\rangle$ into the basis of $(12,34)$ and carry out the measurement of $n^{(f)}_{12}$ in Fig.\,\ref{fig:Fig6-TCM}(d).
Up to an overall phase factor, the final state can be grouped as
\begin{equation}
  |\Psi^{(d)}\rangle = 
 \begin{cases}
     |0\rangle_{12} \big[\alpha|0\rangle_{34} + i\beta|1\rangle_{34} \big], n_{13}=n_{14},n^{(f)}_{12}=0 \\
     |0\rangle_{12} \big[\alpha|0\rangle_{34} -i \beta|1\rangle_{34} \big], n_{13}\ne n_{14},n^{(f)}_{12}=0 \\
    |1\rangle_{12} \big[i\alpha|1\rangle_{34} + \beta|0\rangle_{34} \big], n_{13}=n_{14},n^{(f)}_{12}=1 \\
    |1\rangle_{12} \big[i\alpha|1\rangle_{34} - \beta|0\rangle_{34} \big], n_{13}\ne n_{14},n^{(f)}_{12}=1
    \end{cases}
\end{equation}
A close look at the final state tells us that if we apply an operator $\hat{P}$ as defined below to the final state, it is the result of a braiding operation on the initial state
\begin{equation}
 \hat{P}|\Psi^{(d)}\rangle = \hat{R}_{34} |\Psi^{(a)}\rangle,
 \label{eq:MBB}
\end{equation}
where $\hat{R}_{34}=(1+\gamma_4\gamma_3)/\sqrt{2}$ and $\gamma_{3,4}$ are the Majorana operators of anyons $3$ and $4$.
The operator $\hat{P}$ depends on the measurement outcomes and is given by
\begin{equation}
   \hat{P} = 
 \begin{cases}
     \hat{I}, \qquad \qquad \;\;\; n_{13}=n_{14},n^{(f)}_{12}=0 \\
     \hat{Z}\equiv i\gamma_3\gamma_4, \quad n_{13}\ne n_{14},n^{(f)}_{12}=0 \\
    \hat{Y}\equiv i\gamma_1\gamma_4, \quad n_{13}=n_{14},n^{(f)}_{12}=1 \\
    \hat{X}\equiv i\gamma_1\gamma_3, \quad  n_{13}\ne n_{14},n^{(f)}_{12}=1
    \end{cases}
\end{equation}
Operator $\hat{P}$ has an intuitive interpretation: if we encode information in the parity of anyons $3$ and $4$, then $\hat{P}$ is the single-qubit operator manipulating the degree of freedom associated with anyons $3$ and $4$ as shown in Fig.\,\ref{fig:Fig6-TCM}(e).
By monitoring the measurement outcomes, we apply one of the logical Pauli operators from the set $\{I, \hat{X},\hat{Y},\hat{Z}\}$ to complete the measurement-based braiding in Eq.\,(\ref{eq:MBB}).
In the surface code setting, $\hat{X}$, $\hat{Y}$ and $\hat{Z}$ are the logical operators connecting the twist defects which commute with the stabilizers. 
For example, $\hat{Z}$ operator is defined as a string of Pauli operators highlighted in red circles in Fig.\,\ref{fig:Fig1-JWT} and Fig.\,\ref{fig:Fig4-ParityM}.

We can envision using the MBB to demonstrate the non-Abelian statistics of Majorana fermions.
A minimal set of $6$ twist defects would be enough. In addition to twist $1$ to $4$ in Fig.\,\ref{fig:Fig6-TCM}, we have to include twists $5$ and $6$.
We can initialize in the basis of $(35,46)$ with the vacuum state.
Then, we perform a MBB to braid $3$ and $4$. If we measure the parity of $3$ and $5$: $P_{35}$, there will be $50\%$ probability that the parity is changed.
Generally, we can perform $n$ MBB of $3$ and $4$, and then measure $P_{35}$.
The probability of parity change will be $\frac{1}{2}$, $1$, $\frac{1}{2}$, $0$ for $n=1,2,3,0$ (mod $4$) \cite{HyartPRB13}.
This is an appealing alternative to the condensed matter settings because the necessary experimental capacities are within reach due to recent advancements in the surface code \cite{BarendsNat14,KellyNat15,CorcolesNatComm15}.

\section{Conclusions}
\label{Conclusion}
In conclusion, we have revisited the surface code model with twist defects.
Using a variety of different techniques, we have shown explicitly the emergence of unpaired Majorana zero modes associated with twist defects.
We also propose a scheme to measurement the parity (topological charge) of pairs of Majoranas.
The parity measurement serves as a building block for measurement-based braiding.
We investigate the possibility of performing such braiding without forced measurements.
It turns out that it can be done with a cycle of three topological charge measurements and one additional logical operation.
The uncertainty associated with forced measurements is removed by our approach.
This makes measurement-based braiding an appealing method for both demonstrating non-Abelian statistics of Majorana fermions and building Clifford gates for topological quantum computation.

\section*{Acknowledgments}
We acknowledge support from the ARO, ARL CDQI, ASFOR MURI,
NBRPC (973 program), the Packard Foundation and the Alfred P. Sloan Foundation. 
We thank Steven M. Girvin, Richard T. Brierley, Matti Silveri, Victor V. Albert, Jukka Vayrynen, and Barry Bradlyn for fruitful discussions.

\bibliography{MF_Twist}

\end{document}